\documentclass[preprint]{aastex}

\slugcomment{}

\shorttitle{Dispersion Measurements in Distant Disk Galaxies}
\shortauthors{Davies et al.}

\usepackage{color}

\begin{document}

\title{How well can we Measure the Intrinsic Velocity Dispersion of Distant Disk Galaxies?}


\author{
R.~Davies\altaffilmark{1}, 
N.M.~F\"orster Schreiber\altaffilmark{1},
G.~Cresci\altaffilmark{2},
R.~Genzel\altaffilmark{1,3},
N.~Bouch\'e\altaffilmark{4},
A.~Burkert\altaffilmark{5,1},
P.~Buschkamp\altaffilmark{1},
S.~Genel\altaffilmark{1},
E.~Hicks\altaffilmark{6},
J.~Kurk\altaffilmark{1},
D.~Lutz\altaffilmark{1},
S.~Newman\altaffilmark{7},
K.~Shapiro,\altaffilmark{8}
A.~Sternberg\altaffilmark{9},
L.J.~Tacconi\altaffilmark{1},
S.~Wuyts\altaffilmark{1},
 }
\altaffiltext{1}{Max-Planck-Institut f\"ur extraterrestrische Physik, Postfach 1312, 85741, Garching, Germany}
\altaffiltext{2}{INAF -- Osservatorio Astrofisico di Arcetri, Largo E. Fermi 5, 50125, Firenze, Italy}
\altaffiltext{3}{Department of Physics, Le Conte Hall, University of California, Berkeley, CA 94720, USA}
\altaffiltext{4}{Department of Physics \& Astronomy, University of California, Santa Barbara, USA}
\altaffiltext{5}{Universit\"ats-Sternwarte M\"unchen, Scheinerstrasse 1, 81679 M\"unchen, Germany}
\altaffiltext{6}{Department of Astronomy, University of Washington, Seattle, USA}
\altaffiltext{7}{Department of Astronomy, Campbell Hall, University of California, Berkeley, CA 94720, USA}
\altaffiltext{8}{Aerospace Research Laboratories, Northrop Grumman Aerospeace Systems, Redondo Beach, CA 90278, USA}
\altaffiltext{9}{School of Physics and Astronomy, Tel Aviv University, Tel Aviv, Israel}

\begin{abstract}
The kinematics of distant galaxies, from $z=0.1$ to $z>2$, play a key role in our understanding of galaxy evolution from early times to the present.
One of the important parameters is the intrinsic, or local, velocity dispersion of a galaxy, which allows one to quantify the degree of non-circular motions such as pressure support.
However, this is difficult to measure because the observed dispersion includes the effects of (often severe) beam smearing on the velocity gradient.
Here we investigate four methods of measuring the dispersion that have been used in the literature, to assess their effectiveness at recovering the intrinsic dispersion.
We discuss the biasses inherent in each method, and apply them to model disk galaxies in order to determine which methods yield meaningful quantities, and under what conditions.
All the mean weighted dispersion estimators are affected by (residual) beam smearing.
In contrast, the dispersion recovered by fitting a spatially and spectrally convolved disk model to the data is unbiassed by the beam smearing it is trying to compensate.
Because of this, and because the bias it does exhibit depends only on the signal-to-noise, it can be considered reliable.
However, at very low signal-to-noise, all methods should be used with caution.

\end{abstract}





\section{Introduction}
\label{sec:intro}

Measurements of gas kinematics play a key role in our understanding of the properties and evolutionary state of galaxies \citep{kru78}.
For local galaxies, it is often the deviations from a simple model that lead to new insights.
But in distant galaxies, it is still generally the global velocity and dispersion fields that provide the important physical constraints.
In recent years it has become possible to measure these through the use of integral field spectrometers.
However, a combination of low signal-to-noise and beam smearing due to limited spatial resolution (which increases the observed dispersion while reducing the observed velocity gradient) make it very difficult, in many cases, to extract even these simple quantities reliably.

Some dynamical information can be retrieved without separating velocity and dispersion.
The integrated H$\alpha$ line width implicitly includes both quantities, and has been used to derive the dynamical mass of $z\sim2$ galaxies \citep[e.g.][]{erb06} with an isotropic virial estimator.
Such techniques have also been applied to CO measurements of sub-millimetre and other galaxies at similar redshifts by \cite{tac06,tac08}, who estimated the dynamical mass as an average of virial and global rotating disk estimators.
Similarly, \cite{kas07} avoided the problem when using H$\alpha$ emission to study the Tully-Fisher relation by defining a new parameter that included contributions from both velocity and dispersion.
Dynamical masses can also be derived using spectroastrometry. 
By measuring spatial location as a function of velocity (rather than velocity as a function of spatial position), one can overcome conventional seeing limits by a factor comparable to the ratio between the resolution and the centroid accuracy, and hence probe kinematics to smaller scales \citep{gne10}.
\cite{gne11b} apply this technique to $z\sim2$--3 galaxies.

Integral field spectroscopy, especially when combined with adaptive optics, allows one to perform much deeper analyses. 
Based on spatially resolved kinematics, a number of studies have shown that at $z \sim 1$--3 galaxies can be divided roughly equally into 3 classes, comprising rotating disks, compact dispersion dominated objects, and mergers \citep{for06,for09,sha08,epi09,law09,wri09}.
An important question is how the properties of these early disk galaxies differ from those of local galaxies.
This is an issue that can be addressed in a number of ways using integral field spectroscopy, for example
via evolution of the Tully-Fisher relation \citep{flo06,pue08,vsta08,cre09,epi09,epi10,lem10a,gne11a};
through the turbulence in the disk \citep{gen08,gre10};
using the ratio $V/\sigma$ which measures the importance of angular momentum to the dynamical support \citep{gen06,for06,for09,law07,law09,wri09,pue07,vsta08,lem10b,man11};
and with the Toomre Q parameter \citep{pue10,gen11}.
Similar work has also been performed on lensed galaxies in order to study less luminous, more compact, or more distant galaxies \citep{sta08,jon10}.

One of the key physical issues raised by these studies is the typical local gas velocity dispersion of the disk.
By comparing a sample of nearby galaxies that had been projected to high redshift, to observations of high redshift galaxies, \cite{epi10} showed that the intrinsic dispersions of $z=0$ and $z\sim2$ disk galaxy populations differ significantly.
Indeed, all IFU studies of $z\sim2$ galaxies -- and also $z\sim1$ galaxies \citep{wei06,kas07} -- have found high dispersions to be a unique feature of normal massive disk galaxies at $z>1$, 4--10 times larger than at $z=0$ \citep{dib06}.
This result is also manifest in the thickness of the disks inferred from high resolution images of edge-on galaxies at $z>1$ \citep{elm04}.
These findings suggest that the large dispersions at high redshift are connected with the unique properties of forming galaxies at early cosmic times, such as high gas fractions, global instability to star formation, and clump inspiral by dynamical friction \citep{gen08}.
While feedback from star formation is an obvious candidate for the large high-z dispersions, \cite{gen11} (see also \citealt{aum10}) found only a weak correlation between velocity dispersion and star formation surface density, either in galaxy integrated measurements or as a function of position within disks.
As such, the current observational and theoretical understanding of disk growth at high redshifts via smooth cold gas accretion from the intergalactic medium (e.g. along narrow dense streams) would provide a natural mechanism to generate the elevated dispersions in high redshift galaxies \citep{ker05,ker09,ocv08,dek09}.

However, \cite{gre10} challenge this interpretation, finding that a class of H$\alpha$ bright local galaxies (including mergers) with comparable star formation rates (SFR) and SFR surface densities to $z\sim2$ disks also apparently have comparable ionised gas dispersions.
Since the average gas accretion rate at high redshift is much greater than at low redshift, this raises the question of what it is that maintains the high dispersion.
\cite{gre10} argue that, based on the correlation they find between turbulence and star formation rate, feedback from the star formation itself could be the origin of the disk turbulence at all cosmic epochs.

All the analyses outlined above require a measurement of the intrinsic kinematics of galaxies from the observed kinematics at limited resolution (for galaxies at $z>1$ this is at best 1--2\,kpc with adaptive optics, although more typically $\sim5$\,kpc under natural seeing), and often with limited signal-to-noise;
and specifically to obtain a measure of the mean dispersion that is unbiassed by beam smearing of the velocity gradient.
The efficacy of any particular method depends both on how the kinematics are extracted and also on how the mean (intrinsic) dispersion is then derived from the observed dispersion map.
A number of methods have been used, in particular simple measures of luminosity and uniformly weighted dispersion; a beam smearing correction proposed by \cite{gre10}; and a full disk modelling such as used by \cite{gen08}, \cite{cre09}, \cite{wri09}, and \cite{epi10} to model $H\alpha$ emission, by \cite{gne11a} to model [O{\sc III}], as well as \cite{tac06} to model CO emission.
Our aim in this paper is to compare such commonly used methods; to understand the biasses to which they are subject; and to assess their effectiveness in tracing the intrinsic dispersion and hence the role of pressure support in rotating disks.

\section{Approach}
\label{sec:app}

Technically, the goal of our modelling is to understand quantitatively whether -- and how -- one can reliably correct for the effects of beam smearing in galaxy kinematics in order to infer the intrinsic dispersion, when the spatial resolution and signal-to-noise ratio are limited.
We do this in a very simple way, by creating toy models of disk galaxies that have been convolved with an adopted seeing and to which noise has been added.
We then extract the kinematics from these models using standard tools, and calculate the dispersion using a variety of methods.
By keeping the galaxy models themselves very simple we are able to identify clearly the impact of the beam smearing, without confusion due to other effects.
And by making the entire procedure fully automated, we avoid the possibility of a subjective bias.

To analyse the results we adopt a parameter $\cal P$ that is analogous to a standard deviation.
It measures the RMS difference between the recovered dispersion $\sigma^{out}_i$ and the intrinsic input dispersion $\sigma^{in}$ for the $n$ simulations:
\[
{\cal P} = {\cal S} \left[ \frac{1}{n} \sum^n_{i=1} (\sigma^{out}_i-\sigma^{in})^2 \right]^{1/2}
\]
where 
${\cal S} = \left<\sigma^{out}_i-\sigma^{in}\right> / 
| \left< \sigma^{out}_i-\sigma^{in} \right> |$ is a sign that indicates whether the recovered value is typically more or less than the intrinsic input value.
We use $\cal P$ to measure the quantitative performance of each method used to recover the dispersion.
For the purposes of this paper only, we adopt the following two criteria:
$|{\cal P}| \le 20$\,km\,s$^{-1}$ indicates that $\sigma^{out}$ is (barely) sufficiently accurate to distinguish thick ($\sigma^{in}\sim50$\,km\,s$^{-1}$) and thin ($\sigma^{in}\sim10$\,km\,s$^{-1}$) disks;
and $|{\cal P}| \le 10$\,km\,s$^{-1}$ indicates that $\sigma^{out}$ is essentially `correct'.
In addition to the overall value of $\cal P$, we also look at trends with the size and mass of the galaxies, which would reveal any systematic bias that depends on these properties.
In particular the lowest mass and size bins allow us to consider respectively the two special cases of non-rotating galaxies (because the masses we use refer only to the rotationally supported mass, which is equivalent to the intrinsic velocity gradient) and compact, poorly resolved galaxies.

In the following subsections we describe the disk models we use and the parameter space we cover; the methods used to extract the kinematics and estimate the intrinsic dispersion; and the biasses that are inherent in some of these methods, particularly when working at low signal-to-noise.
In the rest of this Section, we examine these issues in detail for low redshift ($z=0.15$) disks, and then turn to the more challenging case of very low signal-to-noise high redshift ($z=2$) disks in Section~\ref{sec:highz}.

\subsection{Disk Models}
\label{sec:models}

\begin{deluxetable}{lcc}
\tablecolumns{3}
\tablewidth{0pt}
\tablecaption{\label{tab:pars}Ranges of Parameters used in Disk Simulations}
\tablehead{
\colhead{} & \colhead{low redshift}  & \colhead{high redshift} \\
\colhead{} & \colhead{$z \sim 0.15$} & \colhead{$z \sim 2$} }
\startdata
scale                & 2.5\,kpc/\arcsec & 8.0\,kpc/\arcsec \\
seeing               & 1.0\arcsec & 0.5\arcsec \\
mass\tablenotemark{a}& \multicolumn{2}{c}{$(0.1-9)\times10^{10}\,M_\odot$} \\
S\'ersic index         & \multicolumn{2}{c}{$0.5 \le n \le 2.0$} \\
effective radius     & \multicolumn{2}{c}{$2.5 \le R_{eff} \le 5$\,kpc} \\
inclination\tablenotemark{a} & \multicolumn{2}{c}{45$^\circ$} \\
intrinsic dispersion\tablenotemark{b} & \multicolumn{2}{c}{50\,km\,s$^{-1}$} \\
integrated S/N       & 2500 or 40 in 4\arcsec & 20 in 0.8\arcsec \\
\enddata
\tablenotetext{a}{except Fig~\ref{fig:inc} which explores inclinations of 20--80$^\circ$ for a fixed mass of $1\times10^{10}\,M_\odot$.}
\tablenotetext{b}{except Fig~\ref{fig:sn40_sm2_s10}
which uses 10\,km\,s$^{-1}$, and Fig~\ref{fig:sn2500_sm0} which has an additional variable component.}
\end{deluxetable}

We generate disk models using our code DYSMAL, which is described in Appendix~\ref{sec:app:dysmal} and summarised in \cite{cre09}.
For the purposes of this paper, we have modelled the disks as S\'ersic functions with the same luminosity and mass profiles.
The parameters are summarised in Table~\ref{tab:pars}.
We have used a grid of values with S\'ersic index $0.5 \le n \le 2.0$ and effective (i.e. half light) radius $2.5 \le R_{eff} \le 5$\,kpc.
In general we have used a fixed inclination of 45$^\circ$, but a short study of the impact of inclination is given in Section~\ref{sec:inc}.
In order to keep the models simple, we have not included a bulge component.
In fact we do not expect a bulge to significantly change our results because, observationally, kinematics are usually derived from H$\alpha$ or another gas emission line that traces star formation in a disk.
Thus the only impact of a bulge would be to deepen the gravitational potential, and hence modify the intrinsic emission line rotation curve.
We have also assumed that the disks are thin, but have a uniform intrinsic dispersion.
Although it is isotropic in our models, that assumption has no impact on our conclusions. 
We have performed one set of simulations using $\sigma=10$\,km\,s$^{-1}$ which is typical for low redshift disks \citep{dib06}, but for all other cases we 
have fixed its value at 50\,km\,s$^{-1}$.
This is sufficiently greater than zero to allow one to see if the dispersion derived from the observations is an underestimate; but sufficently low that beam smearing of the velocity field can still play a significant role.
We have also performed a second set of simulations using an additional dispersion term defined by $V/\sigma = R/H$.
However, this introduces an extra complication in interpreting the resulting trends, which we discuss in Section~\ref{sec:bias}.
Because of this, we do not include such a term in the main simulations.

The additional parameters we have adopted are:
a seeing of 1.0\arcsec, a spatial sampling of 0.5\arcsec, instrumental broadening of 35\,km\,s$^{-1}$ FWHM, a distance scale for which 1\arcsec\ = 2.5\,kpc (corresponding to $z\sim0.15$), and a rotationally supported mass in the range (0.1--9)$\times10^{10}$\,M$_\odot$.
These are suited to the case of relatively nearby galaxies that is explored in this section.
However, we note that they are not representative of the near-infrared integral field observations of $z>1$ galaxies.
The appropriate parameters for such cases are given in Section~\ref{sec:highz} where we address high redshift galaxies specifically.

Finally, we have created models at high and low signal-to-noise.
These correspond to a S/N of 2500 and 40 in the total line flux integrated within a 4\arcsec\ diameter aperture.
The first case enables us to assess the extraction methods in the ideal case; the latter to understand how their performance changes in a more realistic situation.
For each set of input parameters, we have made disk models with 5 different noise realisations, leading to a total of 225 simulated disks at each S/N level.
When measuring a mean dispersion from the noisy models, occasionally one of the values would be wildly deviant.
We needed to implement a simple way to be robust against such events since the entire process is automated, and so we reject the lowest and highest values among those derived for each set of input parameters.


\subsection{Extracting the Intrinsic Dispersion}
\label{sec:extract}

The methods we use to derive a measure of the mean dispersion are described below, together with the way in which the kinematics are extracted from the datacube in each case.
In the first 3 methods, this is achieved by fitting a Gaussian to the observed line-of-sight velocity distribution;
and the instrumental broadening is subtracted afterwards using a quadrature correction as 
$\sigma_{corr}^2 = \sigma_{obs}^2 - \sigma_{instr}^2$.
In the fourth case, the instrumental broadening is already accounted for during the extraction process.

\begin{enumerate}

\item
The dispersion at every spatial location is measured by fitting a Gaussian to the observed line profile, and the instrumental broadening subtracted in quadrature. From the resulting corrected dispersion map, one calculates a luminosity weighted mean dispersion. For consistency with the nomenclature of \cite{gre10} we call this $\sigma_m$.

\item
The same initial steps are performed as above. The only difference is that at the end one calculates a uniformly weighted mean dispersion, which we denote $\sigma_{m,uniform}$.

\item
A map of the observed dispersion is extracted as previously by fitting a Gaussian to each line profile, but an attempt is made to correct for the beam smearing following the prescription given in \cite{gre10}.
The observed velocity field is re-sampled to a finer grid and used together with the fixed instrumental broadening to create a model of a disk that is then smoothed according to the seeing. 
This yields a spatially variable estimate of the dispersion due to beam smearing, which is subtracted in quadrature from the extracted dispersion map.
A luminosity weighted mean of the resulting values is then calculated. Following \cite{gre10} we call this $\sigma_{m,corr}$.

\item
The final method follows the procedure used by \cite{cre09}.
The kinematics are extracted using our code LINEFIT, described in  Appendix~\ref{sec:app:linefit} (see also \citealt{for09}).
This already takes the instrumental broadening into account, so no additional quadrature correction is needed.
The flux and kinematics maps it produces, together with their uncertainties, are fed directly into our code DISKFIT \citep{cre09} which uses a genetic algorithm to minimise the difference between an exponential disk model and the observations. We note that the radial profile is, following \cite{cre09}, fixed to be exponential even though we use a range of S\'ersic indices when generating the disk models (see Sec.~\ref{sec:models}). 
This will inevitably affect the performance of the method. 
The reason for imposing such a restriction is simply that there is often insufficient signal-to-noise to derive it from the data.
For similar reasons, and also because the clumpiness of high redshift disks means they do not necessarily have line or continuum emission peaking in the centre, the centre needs to be fixed manually \citep{cre09}. 
We have simply set it at the correct location.
The routine first derives $R_{eff}$ from the curve of growth of the luminosity profile. 
It then performs a minimisation on the kinematics to find the best position angle, inclination, systemic velocity, mass, and uniform intrinsic dispersion. 
This last quantity is denoted $\sigma_{diskfit}$.

\end{enumerate}

\subsection{Biasses}
\label{sec:bias}

By definition, the luminosity weighted mean $\sigma_m$ is biassed towards brighter regions (and at low signal-to-noise, so too is the uniformly weighted mean).
Because these are typically closer to the centre where the intrinsic rotation curve has a steeper gradient, this estimator inevitably strongly affected by beam smearing.
During the modelling and fitting iterations, it became clear that some of the methods described in Section~\ref{sec:extract} are subject to less obvious biasses that have a significant impact on the derived mean dispersions particularly when the signal-to-noise is low.

The first arises from performing a subtraction in quadrature,
$\sigma_{corr}^2 = \sigma_{obs}^2 - \sigma_{instr}^2$, when correcting for the instrumental broadening.
With noisy data, there is a chance that $\sigma_{obs}^2 < \sigma_{instr}^2$ which would lead to a negative value for $\sigma_{corr}^2$.
Spatial locations where this occurs must be rejected when calculating the mean dispersion.
However, this imposes a strong bias to high values of $\sigma_{corr}$ since it is more likely to occur at locations where $\sigma_{obs}$ is small.
This noise bias is a critical issue for all methods that involve a quadrature correction; and is the primary reason for the shift, when the signal-to-noise is low, towards higher mean dispersion for the first 3 methods described in Section~\ref{sec:extract}.
Although we have not pursued alternative ways that might deal with this effect, three possible methods to ameliorate its strength could be:
(i) set affected pixels to zero rather than rejecting them,  
(ii) subtract the instrumental broadening in quadrature as the final step, after calculating the mean dispersion, or
(iii) bin the data, for example with a Voronoi tessellation, to reach a minimum signal-to-noise level in each bin.

\begin{figure}
\epsscale{0.7}
\plotone{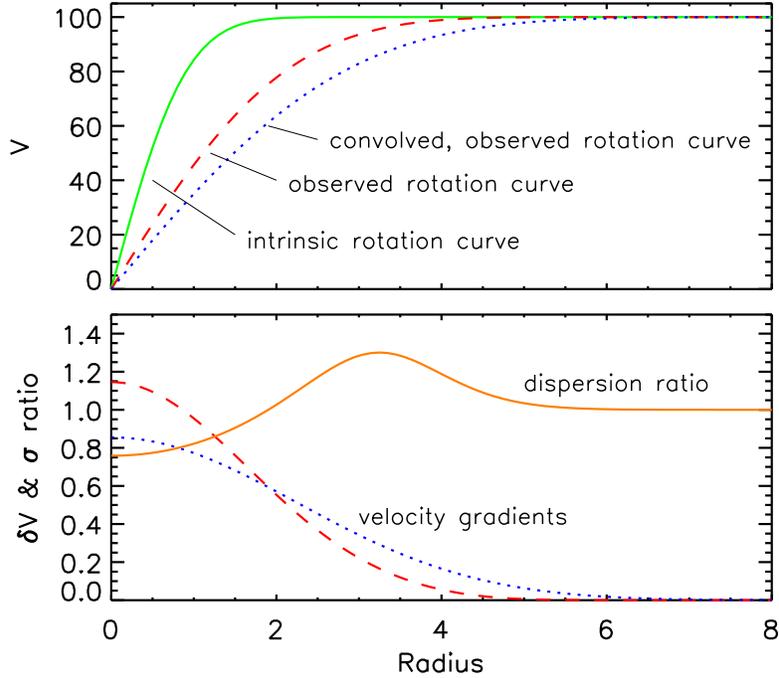}
\caption{\label{fig:rotcurve}
Illustration of how using the observed velocity gradient to infer beam smearing (method~3 in Sec.~\ref{sec:extract}) can lead to both under- and over-correction of its effects.
Upper panel: the central gradient of the intrinsic rotation curve (solid line) is reduced by beam smearing to produce the observed rotation curve (dashed line), but at the same time the radial extent over which a velocity gradient is measured increases.
The same happens again if the observed velocity field is convolved with the seeing to produce the `convolved observed rotation curve'.
Lower panel: the resulting velocity gradients for the latter 2 rotation curves are shown (dashed and dotted lines).
The ratio of these (solid line) indicates at which radii the gradient $\delta V$ of a more smeared velocity field exceeds that of a less smeared one -- and thus where this method overestimates the dispersion due to beam smearing (dispersion ratio $>1$).
}
\end{figure}

Another issue of which one needs to be cautious is over- and under-correcting for beam smearing effects, which are both inherent in the method of \cite{gre10}.
We illustrate how this occurs in Figure~\ref{fig:rotcurve},
which is a conservative lower limit to the scale of the effect because our toy models do not include a bulge.
When attempting to deduce the impact of seeing, these authors convolve the observed velocity gradient (resampled to a finer grid, but otherwise unchanged) with the seeing.
Thus the intrinsic velocity gradient has been convolved twice.
The impact of this is that the estimate of the dispersion induced by beam smearing is too low at the centre where the velocity gradient is steep, but is too high in a band around this region.
Furthermore, during the subsequent quadrature correction, these pixels are more likely to be rejected for the same reasons as described above, imposing an extra bias towards high dispersion in the resulting mean.

The disk fitting method is also not immune to bias.
Some of the parameters that are being derived are poorly constrained at low signal-to-noise.
The most obvious example is inclination, which is strongly correlated with the derived mass.
Thus, while there are large errors in each of these separately, the product $M\,sin^2i$ is much more robust, and correlates well with the input model values (for $S/N=40$ it has a systematic offset of only 0.1\,dex and a standard deviation of 0.2\,dex across the entire parameter range).
The uncertainties in these parameters therefore have little impact on the recovered intrinsic dispersion.
But errors in other parameters can lead to an overestimate of the velocity gradient and hence more of the measured dispersion being attributed to beam smearing.
As a result, the estimate of the intrinsic dispersion will be lower, an effect that is exacerbated at low signal-to-noise.
It is not fully clear why this happens; certainly 
the detailed impact of mismatches in parameters depends on how they are derived and whether they are determined independently beforehand or as part of the fitting procedure.
To reduce the impact of this effect, it is worth attempting to fix {\it a priori} any parameters (e.g. position angle) that can reasonably be derived independently. 
On the other hand, we note that for the DISKFIT code we use here, the intrinsic dispersion was only rarely grossly overestimated.

\begin{figure*}
\epsscale{0.5}
\plotone{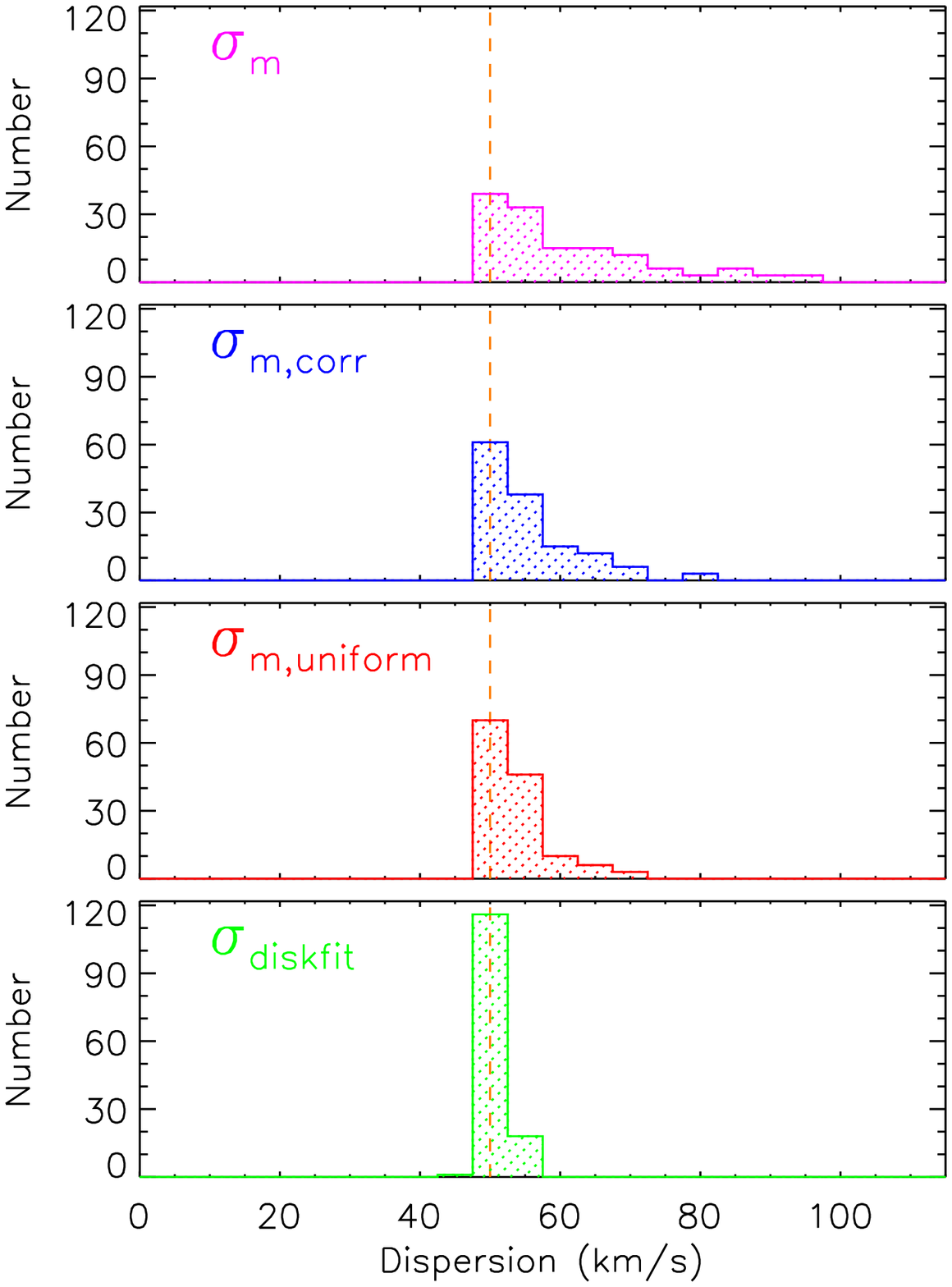}
\caption{\label{fig:sn2500_sm2}
Bar charts showing the distribution of mean dispersions measured in $z\sim0.15$ disks (for details of the models, see Section~\ref{sec:models}) simulated at high signal-to-noise, using the 4 methods described in Sec.~\ref{sec:extract}: 
$\sigma_m$ is a simple luminosity weighted mean;
$\sigma_{m,uniform}$ is a simple uniformly weighted mean;
$\sigma_{m,corr}$ is a luminosity weighted mean after applying the correction derived from the observed velocity field;
$\sigma_{diskfit}$ is the value found by fitting disk models to the simulations.
The dashed line indicates the intrinsic dispersion that was added to the models, and which we are trying to recover.
}
\end{figure*}

\begin{figure*}
\epsscale{0.5}
\plotone{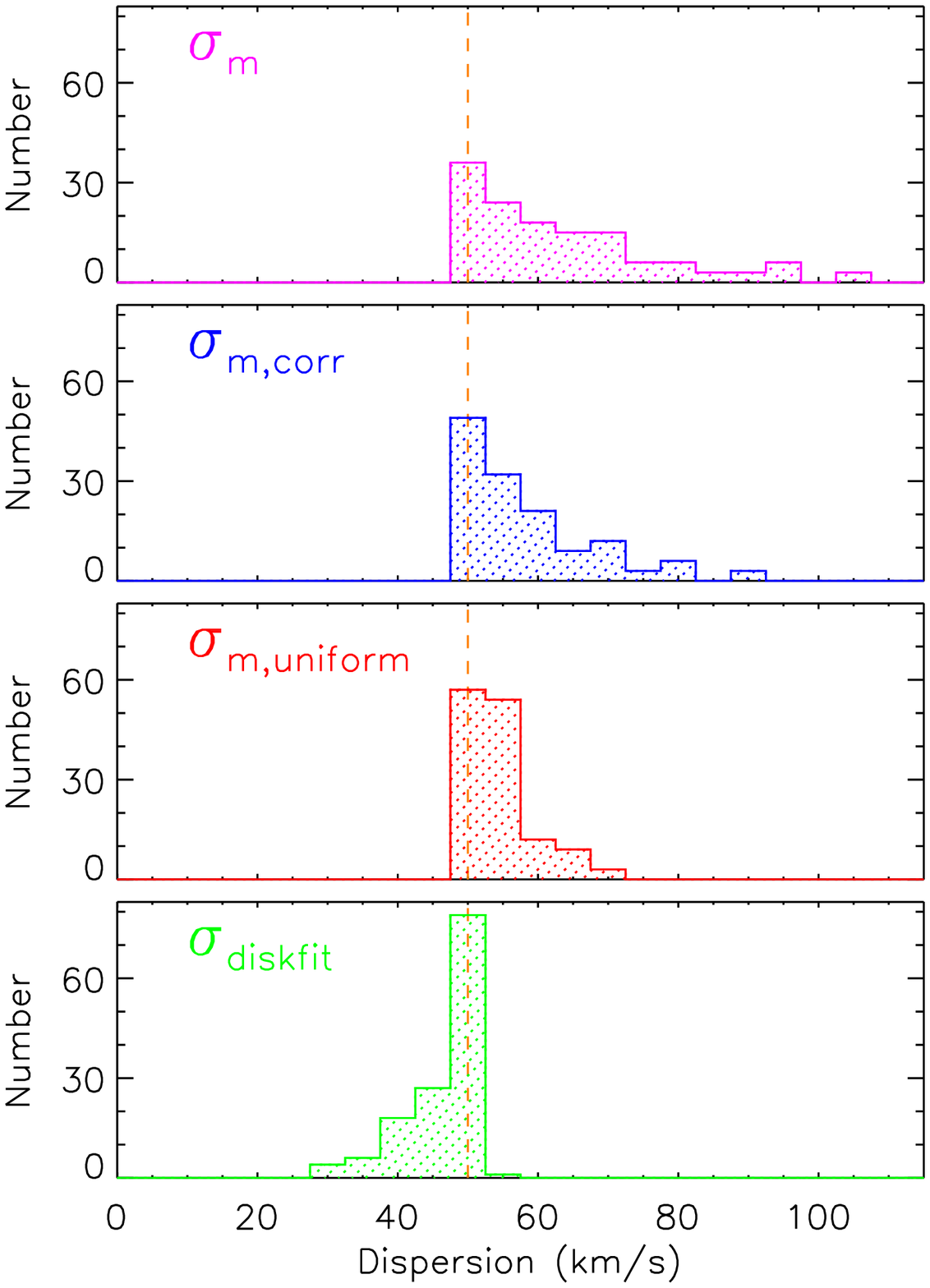}
\caption{\label{fig:sn2500_sm0}
As for Fig.~\ref{fig:sn2500_sm2} but with an additional dispersion term included in the models, defined by $V/\sigma = R/H$ (see Appendix~\ref{sec:app:dysmal}).
Compared to the simulations shown in Fig.~\ref{fig:sn2500_sm2}, only $\sigma_{m,uniform}$ is mostly unaffected.
Both luminosity weighted means $\sigma_m$ and $\sigma_{m,corr}$ are biassed towards higher values, while $\sigma_{diskfit}$ is biassed towards lower values.
}
\end{figure*}

A fourth bias arises from how one models the dispersion in the galaxies.
For the main analysis in this paper we have adopted a uniform intrinsic dispersion across the whole galaxy.
However including an additional dispersion in the model disks such that $V/\sigma = R/H$ (see Appendix~\ref{sec:app:dysmal}) has a noticeable impact on the results.
The effect of this in the models is to increase the dispersion at the centre of the disk where $R$ is small.
Because there is some degeneracy with beam smearing, all the methods used to recover the dispersion are affected.
The impact it has is illustrated by the respective trends seen in Figs.~\ref{fig:sn2500_sm2} and~\ref{fig:sn2500_sm0}, and for which the $\cal P$ values are summarised in Tables~\ref{tab:sn2500_sm2} and~\ref{tab:sn2500_sm0}.
These are for the high signal-to-noise simulations, but similar effects are seen also at low signal-to-noise.
The 3 methods that calculate a weighted mean dispersion have increased values of $\cal P$ when the extra dispersion term is included.
Of these, the uniformly weighted mean $\sigma_{m,uniform}$ shows the least bias simply because it is less influenced by the central regions, and still yields values that are correct (according to the specific definition given in Section~\ref{sec:app}).
Surprisingly, $\sigma_{diskfit}$ from the disk fitting method is biassed slightly downwards, although it too can still be considered correct across the full parameter range.
The reason appears to be that, due to the degeneracy, the code fits part of the extra central dispersion as an increased beam-smeared velocity gradient.
Since this increases the dispersion attributed to beam smearing also at larger radii, the derived value for the uniform intrinsic dispersion term is lower.

\begin{deluxetable}{lrrrrrrrrr}
\tablecolumns{8}
\tablewidth{0pt}
\tablecaption{\label{tab:sn2500_sm2}Values of $\cal P$ (km\,s\,$^{-1}$) corresponding to Fig.~\ref{fig:sn2500_sm2}\tablenotemark{a}}
\tablehead{
\colhead{} & 
\colhead{overall}  & 
\colhead{} & 
\multicolumn{3}{c}{$R_{eff}$} &
\colhead{} & 
\multicolumn{3}{c}{$\log{M/M_\odot}$} 
\\
\cline{4-6} \cline{8-10} \\
\colhead{} & 
\colhead{} & 
\colhead{} & 
\colhead{1.0\arcsec} &
\colhead{1.5\arcsec} &
\colhead{2.0\arcsec} &
\colhead{} & 
\colhead{$ < 9.5$} & 
\colhead{9.5--10.5} & 
\colhead{$ \ge  10.5$}
}
\startdata
$\sigma_m$         & 15.9 && 23.4 & 12.2 & 7.6 && 0.4 & 8.4 & 23.6 \\
$\sigma_{m,corr}$    & 8.7 && 13.3 & 6.0 & 3.4 && 0.2 & 4.7 & 12.9 \\
$\sigma_{m,uniform}$ & 5.6 && 8.6 & 3.7 & 2.1 && 0.2 & 2.9 & 8.3 \\
$\sigma_{diskfit}$   & 1.5 && 1.4 & 1.5 & 1.7 && 0.3 & 1.6 & 1.8 \\
\enddata
\tablenotetext{a}{High S/N $z \sim 0.15$ disks with 50\,km\,s$^{-1}$ intrinsic dispersion.}
\end{deluxetable}

\begin{deluxetable}{lrrrrrrrrrr}
\tablecolumns{8}
\tablewidth{0pt}
\tablecaption{\label{tab:sn2500_sm0}Values of $\cal P$ (km\,s\,$^{-1}$) corresponding to Fig.~\ref{fig:sn2500_sm0}\tablenotemark{a}}
\tablehead{
\colhead{} & 
\colhead{overall}  & 
\colhead{} & 
\multicolumn{3}{c}{$R_{eff}$} &
\colhead{} & 
\multicolumn{3}{c}{$\log{M/M_\odot}$} 
\\
\cline{4-6} \cline{8-10} \\
\colhead{} & 
\colhead{} & 
\colhead{} & 
\colhead{1.0\arcsec} &
\colhead{1.5\arcsec} &
\colhead{2.0\arcsec} &
\colhead{} & 
\colhead{$ < 9.5$} & 
\colhead{9.5--10.5} & 
\colhead{$ \ge  10.5$}
}
\startdata
$\sigma_m$         & 19.2 && 28.5 & 14.6 & 9.1 && 0.8 & 10.9 & 28.4 \\
$\sigma_{m,corr}$    & 12.7 && 19.4 & 8.8 & 5.2 && 0.6 & 7.5 & 18.6 \\
$\sigma_{m,uniform}$ & 6.6 && 10.3 & 4.2 & 2.5 && 0.2 & 3.5 & 9.8 \\
$\sigma_{diskfit}$   & -6.3 && -4.1 & -7.2 & -7.1 && 0.7 & -4.7 & -8.8 \\
\enddata
\tablenotetext{a}{High S/N $z \sim 0.15$ disks with 50\,km\,s$^{-1}$ intrinsic dispersion and an additional dispersion term corresponding to $V/\sigma=R/H$.}
\end{deluxetable}

\subsection{Measured Dispersions}
\label{sec:results}

\begin{figure*}
\epsscale{0.5}
\plotone{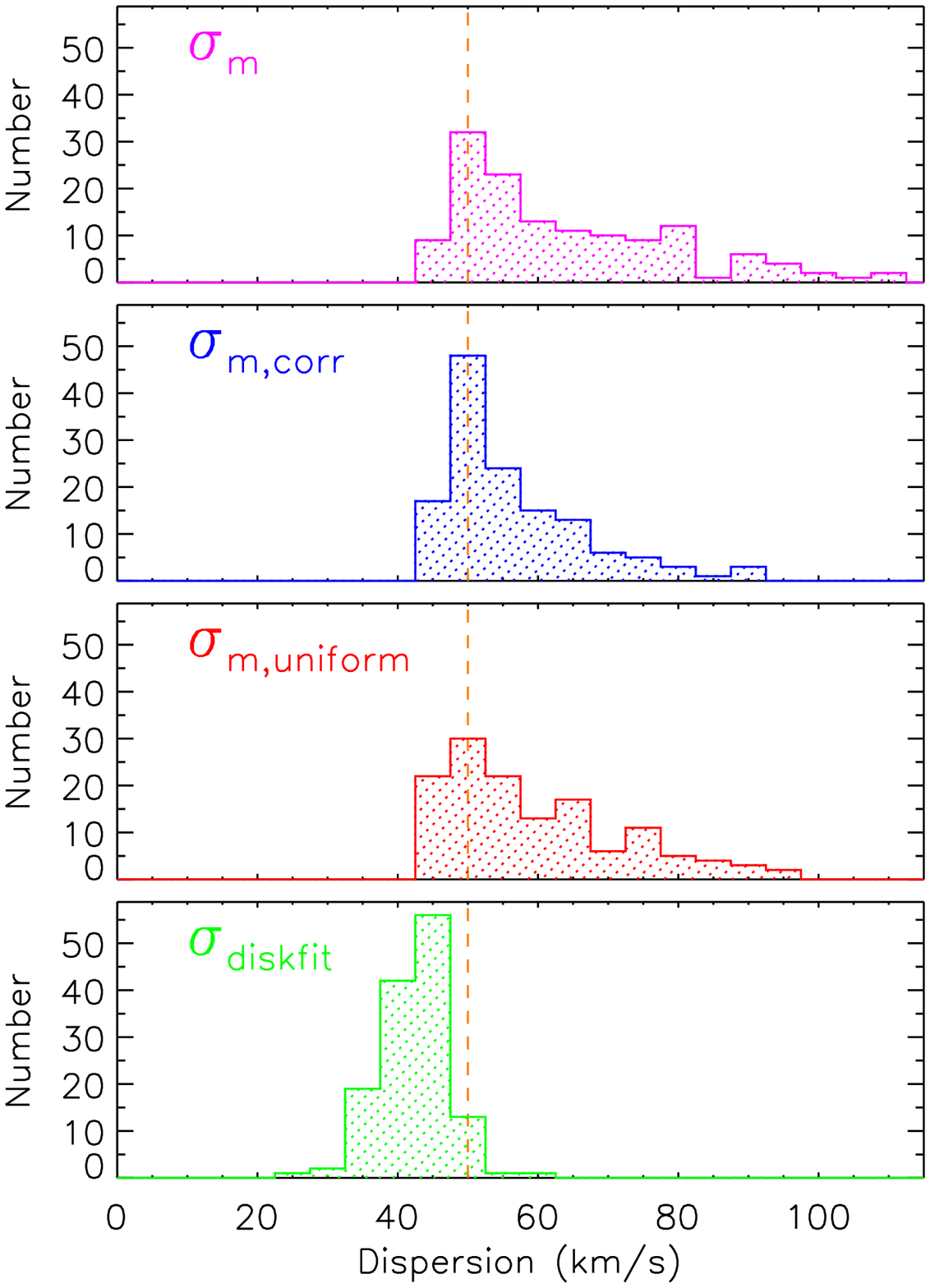}
\caption{\label{fig:sn40_sm2}
As for Fig.~\ref{fig:sn2500_sm2} but the models were simulated at a much lower signal-to-noise, with S/N=40 for the line flux integrated in a 4\arcsec\ diameter aperture.
The uniformly weighted mean $\sigma_{m,uniform}$ performs significantly worse and is now comparable to $\sigma_m$.
Compared to these, $\sigma_{m,corr}$ offers some improvement, although it still shows a tail to higher values suggesting there is some residual beam smearing.
The scatter in $\sigma_{diskfit}$ is also now larger, and it underestimates the dispersion (because at this signal-to-noise level, the fitting procedure assigns too much of the dispersion to beam smearing; see Sec.~\ref{sec:bias}).
However, this bias is independent of the mass and size of the input disk model, and hence independent of beam smearing.
}
\end{figure*}

\begin{figure*}
\epsscale{0.8}
\plotone{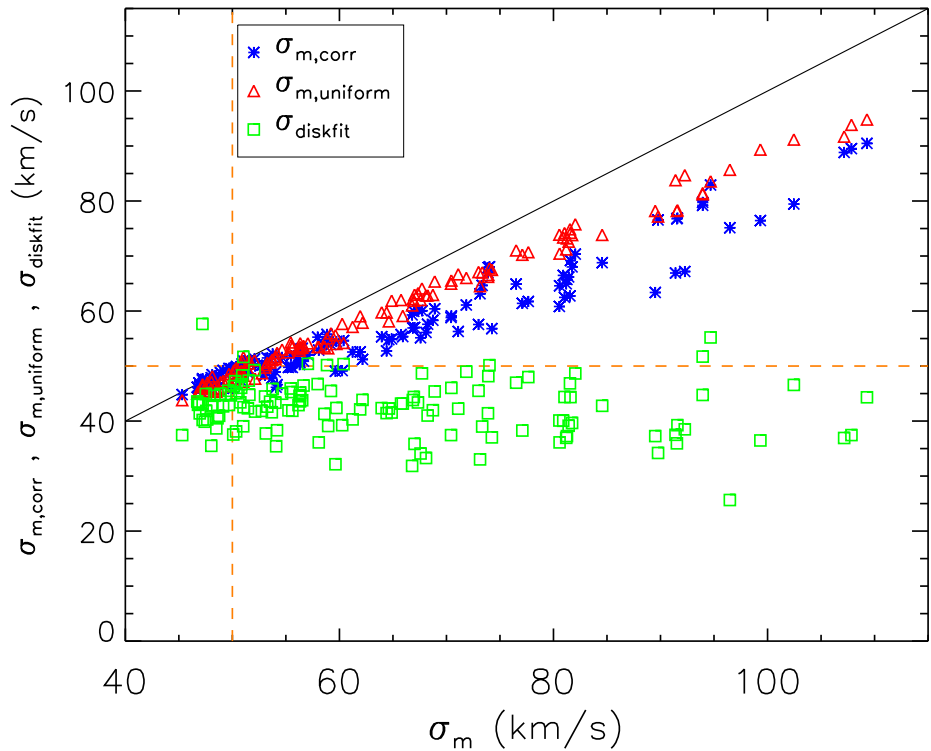}
\caption{\label{fig:sn40_sm2green}
As for Fig.~\ref{fig:sn40_sm2} (models simulated at S/N=40 for the line flux integrated in a 4\arcsec\ diameter aperture) but plotted in a way to match Supplementary Figure~2 of \cite{gre10}.
In this representation the abscissa is $\sigma_m$ (which can be considered as a tracer of beam smearing, see the text in Sec.~\ref{sec:results}).
On this figure, points lying closer to the solid diagonal are more correlated with $\sigma_m$, while those lying closer to the dashed horizontal line represent a better estimate of the intrinsic dispersion.
}
\end{figure*}

\begin{deluxetable}{lrrrrrrrrr}
\tablecolumns{8}
\tablewidth{0pt}
\tablecaption{\label{tab:sn40_sm2}Values of $\cal P$ (km\,s\,$^{-1}$) corresponding to Fig.~\ref{fig:sn40_sm2}\tablenotemark{a}}
\tablehead{
\colhead{} & 
\colhead{overall}  & 
\colhead{} & 
\multicolumn{3}{c}{$R_{eff}$} &
\colhead{} & 
\multicolumn{3}{c}{$\log{M/M_\odot}$} 
\\
\cline{4-6} \cline{8-10} \\
\colhead{} & 
\colhead{} & 
\colhead{} & 
\colhead{1.0\arcsec} &
\colhead{1.5\arcsec} &
\colhead{2.0\arcsec} &
\colhead{} & 
\colhead{$ < 9.5$} & 
\colhead{9.5--10.5} & 
\colhead{$ \ge  10.5$}
}
\startdata
$\sigma_m$         & 21.1 && 29.4 & 17.9 & 12.3 && -2.1 & 10.6 & 31.6 \\
$\sigma_{m,corr}$    & 12.0 && 17.2 & 9.7 & 6.4 && -2.5 & 6.2 & 17.8 \\
$\sigma_{m,uniform}$ & 15.7 && 22.1 & 13.0 & 8.8 && -3.1 & 7.5 & 23.5 \\
$\sigma_{diskfit}$   & -9.0 && -8.4 & -9.0 & -9.5 && -6.9 & -7.9 & -10.7 \\
\enddata
\tablenotetext{a}{S/N=40 disks at $z \sim 0.15$ with 50\,km\,s$^{-1}$ intrinsic dispersion.}
\end{deluxetable}

\begin{figure*}
\epsscale{0.5}
\plotone{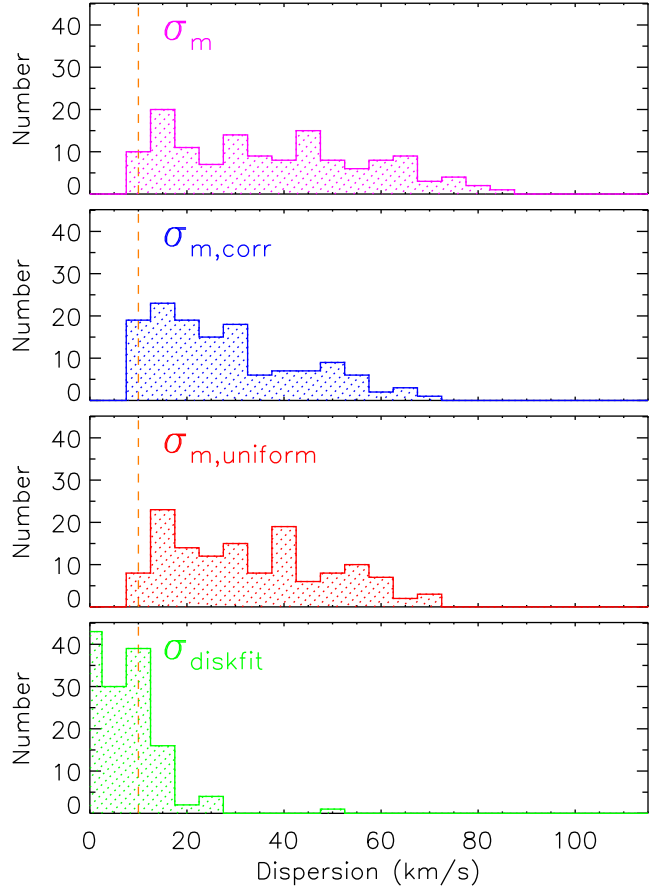}
\caption{\label{fig:sn40_sm2_s10}
As Fig.~\ref{fig:sn40_sm2} but for models with an intrinsic dispersion of 10\,km\,s$^{-1}$ more typical of low redshift galaxies.
}
\end{figure*}

\begin{deluxetable}{lrrrrrrrrr}
\tablecolumns{8}
\tablewidth{0pt}
\tablecaption{\label{tab:sn40_sm2_s10}Values of $\cal P$ (km\,s\,$^{-1}$) corresponding to Fig.~\ref{fig:sn40_sm2_s10}\tablenotemark{a}}
\tablehead{
\colhead{} & 
\colhead{overall}  & 
\colhead{} & 
\multicolumn{3}{c}{$R_{eff}$} &
\colhead{} & 
\multicolumn{3}{c}{$\log{M/M_\odot}$} 
\\
\cline{4-6} \cline{8-10} \\
\colhead{} & 
\colhead{} & 
\colhead{} & 
\colhead{1.0\arcsec} &
\colhead{1.5\arcsec} &
\colhead{2.0\arcsec} &
\colhead{} & 
\colhead{$ < 9.5$} & 
\colhead{9.5--10.5} & 
\colhead{$ \ge  10.5$}
}
\startdata
$\sigma_m$         & 33.8 && 42.2 & 31.9 & 24.9 && 3.2 & 24.0 & 47.6 \\
$\sigma_{m,corr}$    & 23.6 && 31.4 & 20.9 & 15.6 && 2.2 & 15.9 & 33.7 \\
$\sigma_{m,uniform}$ & 28.1 && 35.9 & 25.9 & 20.2 && 3.2 & 19.5 & 39.9 \\
$\sigma_{diskfit}$   & -7.4 && -7.6 & -8.6 & -5.8 && -2.6 & -6.8 & -9.4 \\
\enddata
\tablenotetext{a}{S/N=40 disks at $z \sim 0.15$ with 10\,km\,s$^{-1}$ intrinsic dispersion.}
\end{deluxetable}

Fig.~\ref{fig:sn2500_sm2} shows the results for high signal-to-noise (S/N for the line flux of 2500 in a 4\arcsec\ aperture) disk galaxies at $z\sim0.15$;
and Table~\ref{tab:sn2500_sm2} summarises the calculations of $\cal P$ for them.
This set of simulations represents an ideal case, where the kinematics of a galaxy can be traced out to the faint extended regions at several times the effective radius.
It is because of this that the uniformly weighted mean dispersion $\sigma_{m,uniform}$ performs so much better than the luminosity weighted mean $\sigma_m$.
Indeed, Table~\ref{tab:sn2500_sm2} shows that for $\sigma_m$, $\cal P$ increases rapidly both for more massive and smaller galaxies, suggesting it is strongly correlated with the severity of the beam smearing.
Applying a beam-smearing correction to $\sigma_m$ that is based on the observed velocity field yields $\sigma_{m,corr}$ and brings a clear improvement.
But it too no longer recovers the correct dispersion for the most massive and the smallest disks considered here, suggesting that it has not accounted for all the beam smearing.
On the other hand, in this ideal situation $\sigma_{diskfit}$ does provide a uniformly precise recovery of the intrinsic dispersion across the whole parameter range.

Most observations of distant galaxies are at much lower signal-to-noise.
Fig.~\ref{fig:sn40_sm2} and Table~\ref{tab:sn40_sm2} show a more realistic situation where we have simulated galaxies using the same sets of input parameters but with the S/N of the integrated line flux reduced to $\sim40$.
Fig.~\ref{fig:sn40_sm2green} shows the same data in a different way, with $\sigma_m$ as the abscissa and the other dispersion measures as the ordinate.
In such a representation, points that lie close to the diagonal line are correlated with $\sigma_m$, and hence subject to residual beam smearing effects; and points that are closer to the dashed horizontal line represent a better estimate of the intrinsic dispersion.
It is immediately clear that the uniformly weighted mean is nearly as poor an estimator as the luminosity weighted mean (the slope of the red triangles only just lies under the solid line in Fig.~\ref{fig:sn40_sm2green}).
This is because it is no longer possible to probe the extended fainter parts of the galaxy; in both cases, only the bright central pixels can be measured and this is where the beam smearing is most severe.
Under these conditions, one can achieve a distinct improvement when applying a correction to $\sigma_m$ based on $V_{obs}$: the slope of the blue starred points in Fig.~\ref{fig:sn40_sm2green} is shifted markedly towards the horizontal.
However, the values of $\cal P$ in Table~\ref{tab:sn40_sm2} show that there is still a significant trend with both mass and size of the disk, implying residual beam smearing;
and for the highest masses and smallest sizes the recovered values are barely sufficient to clearly distinguish between thick and thin disks.
While at this S/N level the disk fitting method underestimates the actual dispersion in the model (see Section~\ref{sec:bias}), the dispersion it recovers can still be considered close to the correct value -- although this is the lowest S/N for which it is possible.
In addition, there are only marginal trends with mass and size of the disk model (as also reflected by the horizontal distribution of the green squares in Fig.~\ref{fig:sn40_sm2green}), indicating that while low S/N imposes a bias on the result, the amount of beam smearing does not.
This is an important property, and it means that the disk fitting method can be considered reliable in the sense that (i) bias in the result can be estimated through directly measurable properties of the observation, and (ii) the result is not biassed by the beam smearing it is designed to compensate.

Fig~\ref{fig:sn40_sm2_s10} and Table~\ref{tab:sn40_sm2_s10}
show results for a similar set of simulations for which the intrinsic dispersion in the model was set to 10\,km\,s$^{-1}$, typical of low redshift disks \citep{dib06}.
This lowers the threshold at which dispersion due to beam smearing exceeds the intrinsic dispersion -- in effect making it harder for any of the methods to recover the latter quantity.
The peak in the distributions in Fig~\ref{fig:sn40_sm2}, that was due to the ensemble of galaxies below this threshold for an intrinsic dispersion of 50\,km\,s$^{-1}$, is no longer apparent.
Quantitatively, this means that none of the weighted mean dispersion measures are able to reliably recover the intrinsic dispersion of a partially resolved thin disk: for each of these 3 methods we find overall that ${\cal P} > 20$\,km\,s$^{-1}$, i.e. that typically the intrinsic dispersion is significantly over-estimated.
The only exception is the lowest mass bin which corresponds to a small velocity gradient and hence only moderate beam smearing.
As for the previous simulation, the disk fitting method underestimates the dispersion due to the low signal to noise, but by less than $\sim$10\,km\,s$^{-1}$;
and it exhibits little dependence on disk size or mass.
That its performance is comparable to the previous simulation indicates that at $S/N\sim40$ (spectrally and spatially integrated), $\sigma_{diskfit}$ can be used to reliably discriminate between thin and thick disks, and to recover reasonably accurate estimates of their intrinsic dispersion.

We note that through all the simulations presented here, the performance -- as measured by $\cal P$ -- of all 4 dispersion estimators for the lowest mass disks has been very good.
This is simply because the velocity gradient, and hence beam smearing, in these galaxies is small.
Thus, if one is confident that an observed galaxy is non-rotating, then even the simplest estimator $\sigma_m$ will provide a reliable measure of the dispersion.
On the other hand, for the same reason but in the opposite sense, the 3 weighted mean estimators perform poorly on compact galaxies.
In contrast, to the limits we have explored (where $R_{eff}$ is equal to the seeing and at $S/N\sim40$), the disk fitting method still performs at the same level on the compact galaxies.

Based on these results, we conclude that $\sigma_{m,uniform}$ and $\sigma_{m,corr}$, while performing better than the `beam smearing tracer' $\sigma_m$, are unable to reliably recover the intrinsic dispersion of the disks.
As one might expect, at high signal-to-noise, the faint extended regions provide a reasonable and simple estimate of the intrinsic dispersion. 
It is this that makes $\sigma_{m,uniform}$ effective under such conditions.
On the other hand, it fails at low signal-to-noise because the extended regions cannot be measured.
Instead, $\sigma_{m,corr}$, proposed by \cite{gre10}, provides a better option because it partially corrects for beam smearing in the central regions.
However, because this correction is based on the observed velocity field, it is less effective when the beam smearing contributes substantially to the total observed dispersion.
Thus while $\sigma_{m,corr}$ is sufficient in some situations, it is not an adequate estimator for massive or compact galaxies, or disks with a low intrinsic dispersion.

Iteratively fitting a disk model that is convolved spectrally with the instrumental profile and spatially with the seeing does appear to provide a reliable way to recover the intrinsic mean dispersion of a disk galaxy.
We have found that while there is a tendency for $\sigma_{diskfit}$ to under-estimate the intrinsic dispersion, this bias depends only on the signal to noise (which can easily be measured) and does not adversely impact the results for $S/N \gtrsim 40$.
The method is not strongly biassed by the mass or size of the galaxy -- and hence by the severity of the beam smearing -- and thus is able to distinguish between thick and thin disks.

\subsection{Inclination Effects}
\label{sec:inc}

\begin{figure}
\epsscale{0.8}
\plotone{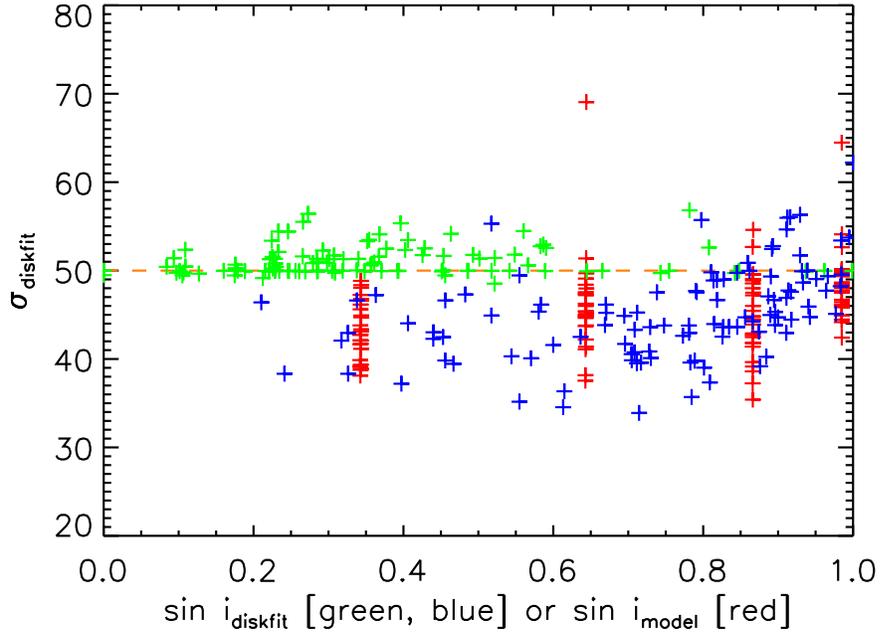}
\caption{\label{fig:inc}
Dispersion recovered using the disk fitting procedure, as a function of inclination (20--80$^\circ$) for $z\sim0.15$ disks at fixed mass ($10^{10}$\,M$_\odot$).
The green points show that there is no dependence on inclination for high signal to noise models (S/N=2500 as for Fig.~\ref{fig:sn2500_sm2}).
The blue points are for low signal-to-noise (S/N=40) models.
They show that, except at the highest sin\,i, the dispersion is underestimated.
This is the same effect as seen in Fig.~\ref{fig:sn40_sm2}, and is due to the low S/N.
In both of these cases, there is a large scatter in the fitted inclination.
The red points therefore show the result for S/N=40 when the inclination is fixed at the correct value: again, at all except the highest sin\,i, the dispersion is underestimated.
}
\end{figure}

There is some evidence that the intrinsic dispersion observed in high redshift disks may depend on inclination, where the more face-on systems systematically have lower dispersion \citep{aum10,gen11}.
While there is only a 2$\sigma$ difference between the mean dispersions of galaxies more and less inclined than $sin(i)=0.64$, confirmation of such an effect would have important implications on our understanding of whether the dispersion is isotropic.
Indeed, the models of \cite{aum10} could at least partially explain this effect.
However, these authors are cautious that, because beam smearing is more severe in more inclined systems, it may not be fully corrected.
In Fig.~\ref{fig:inc} we assess how severe this may be and whether it is likely to contribute to the observed relation.
To do this, we have performed a set of simulations for $z=0.15$ disks at both high and low signal-to-noise for inclinations in the range 20--80$^\circ$, and looked at whether there is any relation between the inclination and dispersion found by the disk fitting method.
At high signal-to-noise, the recovered dispersion is independent of the disk inclination.
This suggests that the disk fitting method itself is robust against such a bias.
At low signal-to-noise, the simulations reveal a possible relation between the two quantities.
Although its 1$\sigma$ level is not really significant, it could, for the parameters we have used, account for as much as half of the trend observed.
This suggests that some caution in interpreting such a relation is warranted, although it seems unlikely that it could account fully for the observations.
We note that in both cases, the scatter in the recovered inclination was large.
However, a third set of simulations where the inclination was fixed during the fitting procedure shows that this scatter is not the source of the trend, and that it is simply a result of fitting a poorly resolved disk at low signal-to-noise.

\section{High Redshift Disks}
\label{sec:highz}

\begin{figure}
\epsscale{0.5}
\plotone{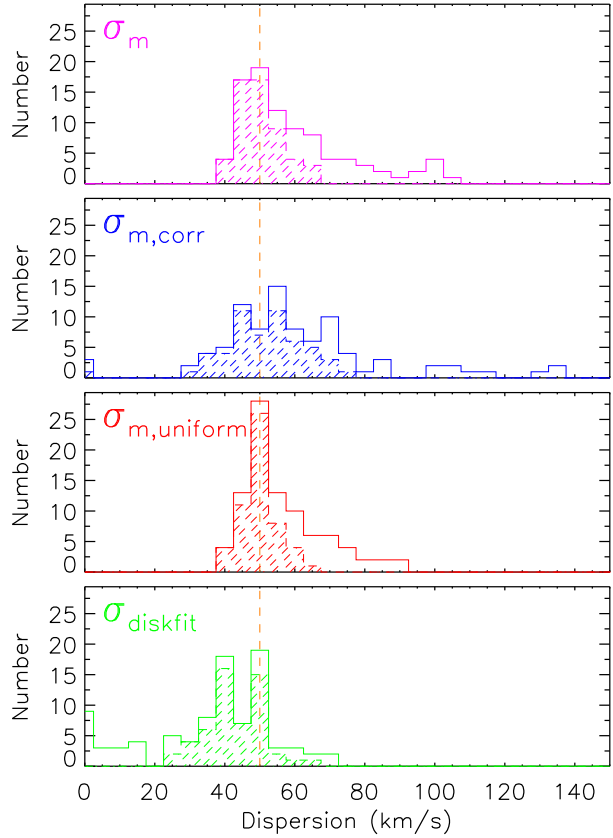}
\caption{\label{fig:sn20_sins}
Mean dispersions for seeing limited observations of $z\sim2$ disks simulated with S/N=20 for the line flux integrated in a 0.8\arcsec\ aperture. 
Note that the scale of the abscissa is different to that in previous figures.
Details of the models are described in Section~\ref{sec:highz}, and the various dispersion measures are as for Fig.~\ref{fig:sn2500_sm2}.
The solid lines show the distribution for all the models;
the shaded regions show the distribution for models with mass $\le 10^{10}$\,M$_\odot$.
In these simulations, none of the methods are able to recover the intrinsic dispersion of the most massive disks, simply because $\sigma_{obs}$ is so completely dominated by beam smearing.
While in most cases this leads to a tail to high values, for the disk fitting method it results in a `zero' measurement.
}
\end{figure}

\begin{deluxetable}{lrrrrrrrrr}
\tablecolumns{10}
\tablewidth{0pt}
\tablecaption{\label{tab:sn20_sins}Values of $\cal P$ (km\,s\,$^{-1}$) corresponding to Fig.~\ref{fig:sn20_sins}\tablenotemark{a}}
\tablehead{
\colhead{} & 
\colhead{overall}  & 
\colhead{} & 
\multicolumn{2}{c}{$R_{eff}$} &
\colhead{} & 
\multicolumn{4}{c}{$\log{M/M_\odot}$} 
\\
\cline{4-5} \cline{7-10} \\
\colhead{} & 
\colhead{} & 
\colhead{} & 
\colhead{0.25\arcsec} &
\colhead{0.6\arcsec} &
\colhead{} & 
\colhead{$ \le 9.5$} & 
\colhead{10.0} & 
\colhead{10.48} & 
\colhead{10.95}
}
\startdata
$\sigma_m$         & 19.0 && 24.8 & 10.2 && -4.5 & 7.9 & 19.6 & 36.2 \\
$\sigma_{m,corr}$    & 26.2 && 31.0 & 20.4 && -13.5 & 9.4 & 25.0 & 48.5 \\
$\sigma_{m,uniform}$ & 12.9 && 16.8 & 7.2 && -4.6 & 6.2 & 15.0 & 23.0 \\
$\sigma_{diskfit}$   & -23.0 && -17.9 & -27.1 && -8.8 & -13.0 & -24.3 & -41.6 \\
\enddata
\tablenotetext{a}{S/N=20 disks at $z \sim 2$ with 50\,km\,s$^{-1}$ intrinsic dispersion.}
\end{deluxetable}

\begin{figure}
\epsscale{0.5}
\plotone{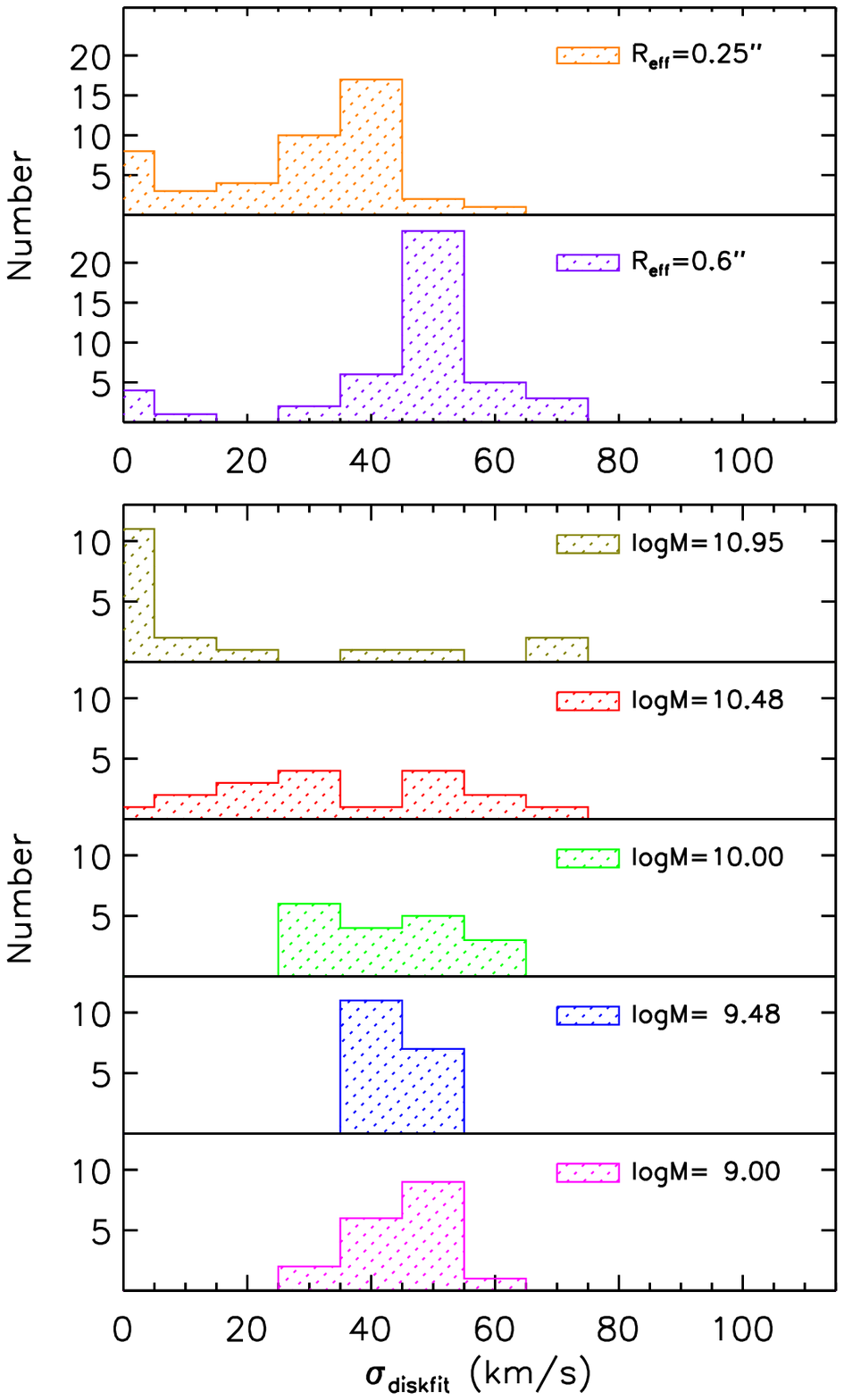}
\caption{\label{fig:pcorbar}
The distributions of $\sigma_{diskfit}$, the dispersion derived from the disk fitting method, displayed for the various input effective radii (top) and for the input mass (bottom).
The method tends to return a low value for compact and massive galaxies.
These are the disks for which beam smearing is most severe, and dominates the total dispersion.
For these models, the disk fitting method attributes most of the dispersion to beam smearing.
As such, it tends not to over-estimate the true local dispersion of the disk.
}
\end{figure}

Observations of disk galaxies at $z\sim2$ are more challenging than those discussed in Section~\ref{sec:app}: the angular scales are smaller and the signal-to-noise is less.
We have therefore performed another set of simulations corresponding to observations such as those reported in \cite{cre09}.
As before, we model disk galaxies (the parameters of which are summarised in Table~\ref{tab:pars}) with a S\'ersic luminosity and mass profile using a range of 
S\'ersic index $0.5 \le n \le 2$ and 
effective radius $2 \le R_{eff} \le 5$\,kpc (i.e. 0.25--0.6\arcsec\ for a scale of $1\arcsec = 8$\,kpc).
This range of sizes has been adopted to match the sample of star forming galaxies studied in \cite{for09} for which $\langle R_{eff} \rangle = 3.1$\,kpc, and in particular the (larger) sizes of galaxies that could be classified as disks using kinemetry \citep{sha08}.
The lower end of our range is consistent with $R_{eff} \sim 2$--3\,kpc of $z\sim2$ main sequence galaxies in the same mass range reported by \cite{wuy11}.
We have adopted an inclination of 45$^\circ$ and an intrinsic dispersion of 50\,km\,s$^{-1}$.
We have also used the same rotationally supported mass range of (0.1--9)$\times 10^{10}$\,M$_\odot$.
We assume that observing conditions are reasonably good, so that the seeing is 0.5\arcsec\ (i.e. without adaptive optics); 
and we have adopted instrumental parameters corresponding to SINFONI in the seeing limited mode (pixel scale 0.125\arcsec\ and instrumental broadening 70\,km\,s$^{-1}$ FWHM, appropriate for the K-band).
Finally, we have set the signal-to-noise for the total line flux integrated in a 0.8\,arcsec\ aperture to be 20, corresponding to the typical S/N in the disks modelled by \cite{cre09}.

The results of the simulations for these parameters are shown in Fig.~\ref{fig:sn20_sins} and Table~\ref{tab:sn20_sins}.
Here, the solid lines reflect the full mass range, while the dashed shaded regions correspond to masses $M \le 10^{10}$\,M$_\odot$.
For masses up to this limit, all the methods appear to work reasonably well and have low values of $\cal P$ (and indeed this makes $\sigma_{m,uniform}$ appear to be the best estimator).
However, as we have seen in Sec.~\ref{sec:results}, this is nothing more than a reflection of the mass (i.e. velocity gradient) at which the intrinsic dispersion is comparable to that induced by beam smearing of the velocity gradient.
If the intrinsic dispersion we had used in the models were lower, this transition mass -- below which all the methods appear to work -- would also decrease.
As shown in Fig.~\ref{fig:sn40_sm2_s10}, this would remove the peak of apparently correctly recovered dispersion values, and extend the tail of over-estimates.

It is straightforward to envisage the improvement that could be achieved using adaptive optics.
Although the only available guide stars (or even tip-tilt stars when using a laser guide star) are often not ideal, it is certainly possible to achieve spatial resolution of 0.1--0.2\arcsec.
This is about 3 times better than the seeing used in the simulations here.
As long as one can integrate long enough to reach sufficient signal-to-noise per pixel (and this is a critical issue if smaller pixels are used so as to better sample the resolution), adaptive optics would have a direct and highly beneficial impact on all the dispersion estimators.

Figure~\ref{fig:sn20_sins} shows that the first 3 methods are characterised by a tail to high values of dispersion.
Table~\ref{tab:sn20_sins} indicates that this tail comprises the higher mass and more compact systems.
This is the same effect that has been seen and discussed in more detail in Sec.~\ref{sec:results}, that the recovered dispersion can still be affected by beam smearing.
As such, since the performance of these methods depends on the amount of beam smearing, one cannot use them to reliably separate beam smearing from intrinsic dispersion.
We note also that at this very low S/N, using the observed velocity gradient to correct for beam smearing has increased the scatter in the measurements.
This is a direct effect of applying a correction that is itself derived from low signal-to-noise data, to low signal-to-noise data.

In contrast to the first 3 methods, the disk fit underestimates the dispersion due to the low signal-to-noise (as discussed in Section~\ref{sec:results}).
Because the S/N is very low, the effect is more severe than in the previous simulations and 
it is notable that in extreme cases, it reverts to a `zero' measurement.
The situations where this can occur are seen in Fig.~\ref{fig:pcorbar} which shows the distribution of $\sigma_{diskfit}$ broken down according to both $R_{eff}$ and Mass.
For larger disks or lower masses, the estimator appears to return a reasonable approximation to the intrinsic dispersion;
but for the highest masses and for more compact disks, there is a broad spread in the recovered values and the dispersion is typically severely underestimated.
As pointed out above, this transition reflects the ratio between the intrinsic dispersion and that induced by beam smearing of the velocity gradient.
It implies that at $S/N\sim20$, $\sigma_{diskfit}$ too is subject to a bias from the severity of the beam smearing and under these conditions should not necessarily be expected to reliably distinguish thick and thin disks in massive systems.

Interestingly, none of the disk fits performed by \cite{cre09} suffered from this effect even though the derived dynamical masses exceeded $10^{10}$\,M$_\odot$.
A possible reason is that their disks are large or have flatter profiles, so that there is more flux in the outer regions:
the effective radius (1.7 times the disk scale length for an exponential profile) from the fits given in their Table~2 is typically 9\,kpc.
This is larger than the more typical values for $z\sim2$ galaxies that we have adopted in our models, and probably results from fitting an exponential profile to the rather flat radial profiles of the galaxies.
As indicated in the top panels of Fig.~\ref{fig:pcorbar}, larger galaxy sizes would tend to reduce the chances of the disk fit failing.
We note that the high value of $\cal P$ for larger disks in Table~\ref{tab:sn20_sins} is due to the impact of the relatively few badly fit cases that have `reverted to zero'.

Nevertheless, the feature of reverting to zero when the fitting fails is an important quality for 2 reasons.
Firstly, in contrast to all the other estimators, $\sigma_{diskfit}$ tends not to over-estimate the intrinsic dispersion.
Secondly, the galaxies for which the disk fit underestimates the dispersion are exactly those for which the other methods overestimate it.
As such, it would apear prudent to use disk fitting together with another estimator to assess whether the intrinsic dispersion has been deduced correctly.
One such possibility would be to look at the ratio $\sigma_{diskfit}/\sigma_{m,uniform}$:
the smaller this ratio, the more likely it is that the dispersion has been incorrectly estimated (although this ratio does not yield any information about what the correct dispersion should be).
An alternative would be to compare $\sigma_{diskfit}$ to the mean dispersion in the outer parts of the galaxy where beam smearing is less severe, as was done by \cite{cre09}.

\section{Conclusions}
\label{sec:conc}

We have analysed a large number of disk galaxies simulated with both high and low signal-to-noise at $z\sim0.15$, and with low signal-to-noise at $z\sim2$.
The aim is to ascertain how reliably one can recover their intrinsic dispersion, and hence how well one can determine the importance of pressure support in the disk.
Technically, the key issue is to correct for beam smearing, which increases the observed dispersion and decreases the observed velocity gradient.
We have compared four different methods of recovering the dispersion using an objective measure of the RMS difference to the input intrinsic dispersion. 
Our conclusions are:
\begin{itemize}

\item
All the methods are subject to biasses, and these need to be understood before interpreting the recovered dispersions. In particular, one should be very cautious of extracting the line-of-sight dispersion by fitting a Gaussian to the line profile and applying a quadrature correction for the instrumental broadening.

\item
Methods that work well on high quality data will not necessarily perform well on noisy data. For example, a simple estimate of the uniformly weighted mean dispersion recovers the intrinsic dispersion well at high signal-to-noise, but fails at low signal-to-noise because it relies on measurements in the extended parts of the disk.

\item
All the mean weighted dispersions are affected by (residual) beam smearing.
Thus, if the intrinsic dispersion is high they may yield a reasonably good estimate; but if the intrinsic dispersion is low they are likely to over-estimate it.
The observed velocity gradient cannot be used to derive the severity of the beam smearing. This is because the more severe the beam smearing itself is, the less the observed velocity gradient.

\item
Fitting a model disk, that has been appropriately spatially and spectrally convolved, to the data is a method that works well although it tends to under-estimate the dispersion at low signal-to-noise.
Despite this, it can be considered reliable because (i) its bias is dependent only on the signal-to-noise and can be easily quantified, and (ii) the result is not biassed by the beam smearing for which it is trying to compensate.

\item
At very low signal-to-noise the disk fitting method returns a `zero' dispersion increasingly often for compact massive disks (because in the fit, all the observed dispersion is attributed to beam smearing).
This behaviour is in contrast to the weighted mean methods which do not fully account for beam smearing in such galaxies and yield an over-estimate of the dispersion.
Thus, when the signal-to-noise is very low, disk fitting and a simpler weighted mean dispersion can be used together to test whether the recovered value is reliable.

\end{itemize}

\acknowledgments

We thank the referee for reading the manuscript and providing useful suggestions.
N.M.F.S. acknowledges support by the Minerva program of the MPG.

\appendix

\section*{APPENDIX}

\section{Modelling Disks with DYSMAL}
\label{sec:app:dysmal}

DYSMAL is a tool that enables one to quantify the impact of spectral and spatial beam smearing on a rotating disk, and so can be used to infer the intrinsic properties of disk galaxies from their observed properties.
As such, it is an observational aid rather than a self-consistent physical model of a galaxy.
It has been used successfully in many cases to interpret observations where a rotating disk is an appropriate model.
These include applying a single dynamical model to observations of different tracers at different resolutions in the centre of NGC\,7469 \citep{dav04a};
assessing the impact of a warped disk on the observed rotation curve in Mkn\,231 \citep{dav04b};
estimating the dynamical mass of some high redshift galaxies by fitting disk models to velocity and dispersion fields \citep{cre09};
revealing residual non-circular motions tracing inflow in the circumnuclear region of NGC\,1097 \citep{dav09};
and in assessing whether or not the line width in dense gas tracers is indicative of a thickened disk in the nuclei of nearby AGN \citep{san11}.

The code first generates a face-on model of a disk galaxy in 3 spatial dimensions [$X_0,Y_0,Z_0$] with the specified radial luminosity profile.
This can be a combination of various functions including Gaussian or Moffat profiles (which can be placed off-axis in order to create rings), as well as power law or S\'ersic profiles.
For each component, one can specify the relative mass weighting and luminosity weighting separately.
This provides the flexibility to see how different tracers (e.g. observations of molecular versus ionised gas) appear for the same underlying gravitational potential.
In addition, one can include a central point mass to represent the influence of a back hole.
The sum of all the components is used to generate a rotation curve under the assumption that the disk is supported entirely by ordered rotational motion.
This is an important issue, and the impact of this assumption is discussed in detail below.
The model is then rotated to the appropriate inclination and position angle to yield a cube, the axes of which are aligned with the sky [$X_s,Y_s,Z_s$].
In parallel, a second matching cube is created that contains information about the line-of-sight velocity at each position, which depends on the derived rotation curve.
The line-of-sight velocity distribution at each projected position [$X_s,Y_s$] is then derived by summing the contributions from each pixel along that sight line $Z_s$ to create a cube that has two spatial dimensions and one velocity dimension [$X_s,Y_s,V$].
During this process one can include the effects of instrumental spectral broadening by applying an additional dispersion term.
Each velocity slice of this cube is then convolved with the spatial beam to yield the final output cube.
In order to allow one to simulate the elliptical beams usually associated with millimetre interferometric observations, the beam in the model need not be symmetrical.
At the end, noise can be added.

The code provides some additional functionality for more specific situations.
Often, a disk is not smoothly illuminated, and so there is an option in the model to apply an intensity mask.
It is also possible to model a warped disk.
This is implemented in a very simple way, splitting the disk into discrete rings, the position angle and inclination of which can be specified individually or as a function of radius.
While this inevitably imprints discontinuities in the resulting model, the beam smearing smoothes over them if the width of the rings is narrow enough.
Finally, the code provides the option of including an expansion velocity, which may be useful when modelling expanding rings.

The dynamical mass is an important issue, particularly when using a thin disk model such as DYSMAL to approximate a thick disk.
The scaling of the rotation curve in DYSMAL is defined by specifying, for some radius, the enclosed mass -- defined as the mass that is supported solely by ordered rotation.
However, many disks depart from the ideal thin disk assumption.
DYSMAL can partially accommodate this by allowing the thickness of each component in the luminosity or mass profile to be defined.
It then optionally includes an estimate of the local dispersion based on the thickness $H$, radius $R$ and velocity $V_R$, or mass surface density $\Sigma_R$.
The calculation can be chosen to be appropriate for either 
a compact disk (i.e. $\sigma = V_R H / R$) or for 
an infinite disk (i.e. $\sigma^2 = 2 \pi G \Sigma_R H$).
The model assumes this dispersion is isotropic.
For non-negligible dispersions or disk thicknesses, some of the mass is supported by random motions.
If the mass of the model galaxy is fixed, the effect of this would be to reduce the rotation velocity.
In the Galaxy this is observed as asymmetric drift, and has been measured as the tendency for populations of stars with higher dispersion to lag increasingly far behind the local standard of rest.
DYSMAL does not explicitly account for asymmetric drift by reducing the rotation velocity when there is a finite dispersion.
Instead, it allows the system mass to increase (and thus the rotation velocity is unaffected).
As such, the user has to be aware that the actual dynamical mass can be greater than that specified, which is only the mass supported by ordered rotation.
If one adopts the simple premise that the dynamical mass is given by 
$M_{dyn} = (V^2 + 3\sigma^2)G/R$ (as adopted by \citealt{dav07}; see for example Appendix B of \citealt{ben92}) then the mass discrepancy will be a factor 2 when $\sigma = V/\sqrt{3}$.

\section{Emission Line Fitting with LINEFIT}
\label{sec:app:linefit}

The proliferation of integral field spectrographs has led to a need
to develop new tools that can extract the required information
from data sets that are often complex to visualise.
The need for such tools is felt particularly in the near-infrared
regime where there are complications due to the bright and variable OH
sky line emission, leading directly to a strong
wavelength dependence in the noise properties of the data.
For very faint sources such as high redshift galaxies, this is
exacerbated by systematic effects.
Spectrally variable noise has a much greater impact than
spatially variable noise (which may arise as a result of dithering the
field of view) because line profiles are typically analysed in each
spatial pixel independently.
And it will have a significant effect on any attempt to measure properties
of emission (or absorption) lines in the science target, especially
those which lie close in wavelength to OH airglow lines.

A code for extracting a parameterisation of line of sight velocity distributions (LOSVDs) has been presented by \cite{Cap04}.
Their emphasis was on how best to perform the fit to obtain robust
estimates of the Gauss-Hermite terms -- $V$, $\sigma$, $h_3$, $h_4$,
etc. -- specifically for stellar absorption features (in SAURON data at
optical wavelengths).
They cope with spatially variable signal-to-noise by penalising the $\chi^2$ when the signal-to-noise is low, which forces the fit to become more Gaussian.
This means that the values derived for the higher order terms
depend on whether each term is allowed to contribute to the fit, which
itself depends on the signal-to-noise at that spatial location.
While this is, in a sense, an optimal extraction of these parameters,
it does lead to difficulties in interpreting and understanding the
results.

We have developed the code LINEFIT that allows one to extract the parameterised LOSVDs of emission lines in a robust way that leads to straightforward interpretation of the output; that can deal with noisy data and a spectrally variable signal-to-noise, both of which are typical of near-infrared data;  and that yields robust estimates of the uncertainties.
Two methods are applied: moment calculations, and profile
fitting by convolution with spectral template.
In the latter case, the template may be either a single line or a
multiplet;
and although in the current implementation the kernel is a single Gaussian, it can be easily generalised to any input profile 
(in this way, one could in principle simultaneously fit H$\alpha$ together with the [N{\sc ii}] lines either side of it, allowing their relative fluxes to vary but keeping their separation fixed).
The main features of the code are that it takes properly into account the
3D noise properties of the input data via weighting in the fits, and it
determines robust and realistic uncertainties on the derived flux and
kinematic parameters using Monte Carlo techniques.
The code has been extensively tested using SINFONI data of nearby
starburst and active galaxies, as well as faint high-redshift sources.

\subsection{Pre-processing}
\label{sec:pre}

The code includes options to perform spatial smoothing by median
filtering each spectral slice, and to extract the line properties only
within a specified region.
But the two main steps of the pre-processing are to resample
the spectral axis of the cube so that it is uniformly sampled in
velocity rather than wavelength or frequency (which can become important when fitting multiple lines simultaneously); 
and to estimate and subtract the continuum.
Both of these are performed on the template profile as well as the
data itself.

Resampling is a necessary step for the template convolution technique, but
also makes the moments calculation simpler.
On the other hand, continuum subtraction is a necessary step before moments
can be calculated.
In principle continuum fitting could be included as part of the
template convolution method.
However, tests have shown that there is little real benefit in doing
so, and a prior estimation is equally as good.
The continuum is estimated from spectral segments either side of the
emission line, specified by the user.
A polynomial (by default a linear function) is then fit iteratively to
these regions.
At each spaxel a spectrally uniform noise is estimated from the
residuals to the fit, and this is used to reject pixels from the
subsequent iteration. 
This estimate of the noise has no further use if there is a valid noise
 cube.
However, if no noise cube has been specified, this estimate of the noise is
also used both to reject deviant pixels during the template
convolution fitting, and as the value of $\sigma$ in the threshold
applied during the moments calculation.

\subsection{Line profile fitting with template convolution}

Perhaps one of the most simple, and commonly used, methods of deriving
the flux, velocity and dispersion of an emission line is to fit a
Gaussian.
Convolving a template of an unresolved line profile with a Gaussian
kernel is only one step more sophisticated, but has two important
advantages.
The first of these is that, as long as the spectral profile is
properly sampled, high frequency noise has much less influence on the
derived properties.
The second is that the resulting dispersion is the intrinsic
dispersion, and has had the instrumental broadening removed via the
convolution.
This is a much more robust way of correcting the dispersion for
instrumental effects than applying a quadrature correction afterwards,
and does not implicitly assume that the instrumental profile is
Gaussian. 

For near-infrared spectroscopy, an unresolved line profile can be generated either using the strong OH emission lines which dominate the background in typical science exposures (although some care is needed to ensure that the line chosen is not adversely affected by blending), 
or an arc line from the wavelength calibration frames.
It is important only that the unresolved line is observed with the
same instrumental setup as the science target (in the case of SINFONI,
pixel scale and grating).
The template line profile does not need to centered in its spectral
segment, since a zero-point offset calibration is performed internally
in the code.

In the most basic implementation which is used here, the kernel is a
Gaussian.
However, it would be straightforward to extend this to more complex or
more detailed profiles.
A $\chi^2$ minimisation is then performed to optimise the kernel for
each spaxel independently where
\[
\chi^2 = \sum_v \{[I_v - (P \otimes K)_v] w_v\}^{2}
\]
where $K$ is the kernel, $P$ is the template profile, $I_v$ is the
measured intensity, and the sum is evaluated over all velocity steps $v$.
During the minimisation, each pixel along the spectral segment is
weighted by $w_v$.
The weights come either from a separate noise cube, if one is supplied, or from the noise estimate derived when fitting the continuum.
The uncertainties in this and the other moments are derived as described in Sec.~\ref{sec:extend}.

When interpreting the output from this template convolution technique,
it is important to understand what has been measured.
If the line profile being quantified is well approximated by a
Gaussian convolved with the instrumental profile, then the fitted
parameters accurately describe the intrinsic properties of the line.
However, this may not be the case if, for example, the line comprises
several components blended together.
In such situations, the properties are likely to reflect the dominant
component of the emission, if one exists.
Alternatively, a large dispersion may reflect that two or more
different components are comparably strong, even though each component
may itself be narrow.
If there are two fully distinct components, it is possible that only one
will be included in the fit.
The effect of this leads to discontinuities in the velocity field
between adjacent spaxels, which would indicate 
that a more complex fit involving multiple components is warranted.

\subsection{Moment Calculations}

The low order moments are well known and have been used for a long
time in radio astronomy, primarily because they yield a meaningful
parameterisation even for complex LOSVDs which, as discussed
above, is the major disadvantage of the template convolution technique.
For simple profiles which are reasonably well approximated by a Gaussian
function, moments are quantitatively similar to that technique (except for the dispersion which, for the moment calculation, is the observed value still including instrumental broadening).
In these cases, it should be noted that the template convolution technique can perform better for data that have low signal-to-noise and relatively poor spectral resolution (for example, as typified by SINFONI observations of high redshift galaxies).
On the other hand, as explained later in this section there are
important differences for most complex profiles.

Here, the standard expressions for the moments are generalised to
allow for variable uncertainties, and evaluated 
within the specified wavelength (velocity) range.
The only tricky issue concerns the thresholding level.
A threshold, below which pixel values are ignored, is necessary to
avoid regions with very low signal imposing a strong bias on the
moments.
Thresholding is a technique widely used in adaptive optics wavefront sensing,
and significantly improves the centroid estimation.
Here it is applied to all the moment calculations.
Extensive simulations suggest that a threshold of $0.5\sigma$ (which
in regions of no signal would remove all the negative pixels and
$\sim40$\% of the positive pixels) is
typically appropriate, but this can be varied by the user.

The flux at each spaxel is given by the zeroth order moment $M_0$,
which is simply the  sum of intensity values across a specified
spectral segment 
\[
M_0 = \Delta v \ \frac{\sum I_v w_v}
                      {\sum w_v}
\]
where $\Delta v$ is the velocity step, $I_v$ is the intensity as a
function of velocity, and
$w_v$ is the weighting applied to each intensity measurement $I_v$.
The uncertainies in this and the other moments are derived as described in Sec.~\ref{sec:extend}.

The first order moment $M_1$ is the centroid of the emission line
distribution
\[
M_1 = \frac{\sum I_v w_v v}
           {\sum I_v w_v}
\]
It is important to realise that this is an intensity weighted mean
velocity (where the intensity itself is also weighted).
For line profiles with asymmetric or multiple components, this will typically yield
a different result to a Gaussian convolution.
Which of the methods should be used depends on whether the aim is to measure an average velocity, or the velocity of the dominant component.
For example, when an emission line has a wing, the template convolution method will return a velocity close to that of the line core, while the first order moment will be shifted towards the wing.
If there are two distinct emission lines, the first order moment could lie between them, while the template convolution will yield the velocity of the stronger line.

Finally, the second order moment $M_2$ is a measure of the
velocity dispersion. It is the intensity weighted root-mean-square
width of the distribution.
For the generalisation in which the intensity is also weighted, it is
defined as
\[
M_2 = \left[ \frac{\sum I_v w_v (v-M_1)^2}
                  {\sum I_v w_v}          \right]^{1/2}
\]
One needs to be careful when interpreting the value of $M_2$ as a dispersion.
For a Gaussian distribution, they are exactly the same.
However, for a top-hat (or any other centrally concentrated) profile,
$M_2$ is smaller than the width of a Gaussian fit to the profile;
similarly, $M_2$ quickly becomes very large if there are broad wings
or other extended components in the profile.

\subsection{Robust estimation of uncertainties}
\label{sec:extend}

For both of methods described above, uncertainties are estimated using the Monte Carlo technique.
In the case of the moments, the uncertainties can also be calculated directly, yielding quantitatively similar results.
For the Monte Carlo method, the code generates 100 realisations of the data.
Each time the value of each individual pixel is perturbed from its original value according to the measured noise statistics.
The noise is taken from the noise cube, if one is supplied.
In this case, it is spatially and spectrally variable.
Alternatively, if no noise cube is available, it can be estimated from the fit to the continuum (see Sec.~\ref{sec:pre}).
In this case, the noise is spatially variable, but spectrally uniform at each spatial position.
The emission line properties are calculated or fitted for each new realisation of the data in exactly the same way as the original data.
The uncertainties are then derived from the distribution of values for each parameter.
As such, they are the uncertainties for joint variation of the emission line parameters.
Both because of this, and because they take into account spectral as well as spatial variations in the noise, they can be considered robust.

\clearpage



\begin{thebibliography}{}

\bibitem[Aumer et al.(2010)]{aum10}
Aumer M., Burkert A., Johansson P., Genzel R., 2010,
ApJ, 719, 1230

\bibitem[Bender et al.(1992)]{ben92}
Bender R., Burstein D., Faber S., 1992,
ApJ, 399, 462

\bibitem[Cappellari \& Emsellem(2004)]{Cap04}
Cappellari M., Emsellem E., 2004,
PASP, 116, 138

\bibitem[Cresci et al.(2009)]{cre09}
Cresci G., et al., 2009,
ApJ, 697, 115

\bibitem[Davies et al.(2004a)]{dav04a} 
Davies R., Tacconi L., Genzel R., 2004a,
ApJ, 602, 148

\bibitem[Davies et al.(2004b)]{dav04b} 
Davies R., Tacconi L., Genzel R., 2004b,
ApJ, 613, 781

\bibitem[Davies et al.(2007)]{dav07}
Davies R., M\"uller S\'anchez F., Genzel R., Tacconi L., Hicks E., 
Friedrich S., Sternberg A., 2007,
ApJ, 671, 1388

\bibitem[Davies et al.(2009)]{dav09}
Davies R., Maciejewski W., Hicks E., Tacconi L., Genzel R., Engel H., 2009,
ApJ, 702, 114

\bibitem[Dekel et al.(2009)]{dek09}
Dekel A., et al., 2009,
Nature, 457, 451

\bibitem[Dib et al.(2006)]{dib06}
Dib S., Bell E., Burkert A., 2006,
ApJ, 638, 797

\bibitem[Elmegreen et al.(2004)]{elm04}
Elmegreen D., Elmegreen B., Sheets C., 2004, 
ApJ, 603, 74

\bibitem[Erb et al.(2006)]{erb06}
Erb D., Steidel, C., Shapley A., Pettini M., Reddy N., Adelberger K., 2006,
ApJ, 646, 107

\bibitem[Epinat et al.(2009)]{epi09}
Epinat B., et al., 2009,
A\&A, 504, 789

\bibitem[Epinat et al.(2010)]{epi10}
Epinat B., Amram P., Balkowski C., Marcelin M., 2010
MNRAS, 401, 2113

\bibitem[Flores et al.(2006)]{flo06}
Flores H., Hammer F., Puech M., Amram P., Balkowski C., 2006,
A\&A, 455, 107

\bibitem[F\"orster Schreiber et al.(2006)]{for06}
F\"orster Schreiber N., et al., 2006, 
ApJ, 645, 1062

\bibitem[F\"orster Schreiber et al.(2009)]{for09}
F\"orster Schreiber N., et al., 2009, 
ApJ, 706, 1364

\bibitem[F\"orster Schreiber et al.(2011)]{for11}
F\"orster Schreiber N., et al., 2011, 
ApJ, accepted

\bibitem[Genzel et al.(2006)]{gen06}
Genzel R., et al., 2008,
Nature, 442, 786

\bibitem[Genzel et al.(2008)]{gen08}
Genzel R., et al., 2008,
ApJ, 687, 59

\bibitem[Genzel et al.(2011)]{gen11}
Genzel R., et al., 2011
ApJ, accepted

\bibitem[Gnerucci et al.(2010)]{gne10}
Gnerucci A., Marconi A., Capetti A., Axon D., Robinson A., 2010,
A\&A, 511, A19

\bibitem[Gnerucci et al.(2011a)]{gne11a}
Gnerucci A., et al., 2011a,
A\&A, 528, 88

\bibitem[Gnerucci et al.(2011b)]{gne11b}
Gnerucci A., et al., 
A\&A, submitted

\bibitem[Green et al.(2010)]{gre10}
Green A., et al., 2010, 
Nature, 467, 684

\bibitem[Jones et al.(2010)]{jon10}
Jones T., Swinkbank A.M., Ellis R., Richard J., Stark D., 2010,
MNRAS, 404, 1247

\bibitem[Kassin et al.(2007)]{kas07}
Kassin S., et al., 2007,
ApJ, 660, L35

\bibitem[Kere\v s et al.(2005)]{ker05}
Kere\v s D., Katz N., Weinberg D., Dav\'e R., 2005,
MNRAS, 363, 2

\bibitem[Kere\v s et al.(2009)]{ker09}
Kere\v s D., Katz N., Fardal M., Dav\'e R., Weinberg D., 2009,
MNRAS, 395, 160

\bibitem[Law et al.(2007)]{law07}
Law D., Steidel C., Erb D., Larkin J., Pettini M., Shapley A., Wright S., 2007,
ApJ, 669, 929

\bibitem[Law et al.(2009)]{law09}
Law D., Steidel C., Erb D., Larkin J., Pettini M., Shapley A., Wright S., 2009,
ApJ, 697, 2057

\bibitem[Lemoine-Busserolle et al.(2010a)]{lem10a}
Lemoine-Busserolle M., Lamareille F., 2010a,
MNRAS, 402, 2291

\bibitem[Lemoine-Busserolle et al.(2010b)]{lem10b}
Lemoine-Busserolle M., Bunker A., Lamareille F., Kissler-Patig M., 2010b,
MNRAS, 401, 1657

\bibitem[Mancini et al.(2011)]{man11}
Mancini C., et al., 2011,
in prep

\bibitem[Ocvirk et al.(2008)]{ocv08}
Ocvirk P., Pichon C., Teyssier R., 2008,
MNRAS, 390, 1326

\bibitem[Puech et al.(2007)]{pue07}
Puech M., Hammer F., Lehnert M., Flores H., 2007,
A\&A, 466, 83

\bibitem[Puech et al.(2008)]{pue08}
Puech M., et al., 2008,
A\&A, 484, 173

\bibitem[Puech(2010)]{pue10}
Puech M., 2010,
MNRAS, 406, 535

\bibitem[Sani et al.(2011)]{san11}
Sani E., et al., in prep.

\bibitem[Shapiro et al.(2008)]{sha08}
Shapiro K., et al., 2008,
ApJ, 682, 231

\bibitem[Stark et al.(2008)]{sta08}
Stark D., Swinkbank A.M., Ellis R., Dye S., Smail I., Richard J., 2008,
Nature, 455, 775

\bibitem[Tacconi et al.(2006)]{tac06}
Tacconi L., et al., 2006,
ApJ, 640, 228

\bibitem[Tacconi et al.(2008)]{tac08}
Tacconi L., et al., 2008,
ApJ, 680, 246

\bibitem[van der Kruit \& Allen(1978)]{kru78}
ven der Kruit P., Allen R., 1978,
ARA\&A, 16, 103

\bibitem[van Starkenburg et al.(2008)]{vsta08}
van Starkenburg L., van der Werf P., Franx M., Labb\'e I., Rudnick G., 
Wuyts S., 2008,
A\&A, 488, 99


\bibitem[Weiner et al.(2006)]{wei06}
Weiner B., et al., 2006,
ApJ,  653, 1027

\bibitem[Wright et al.(2009)]{wri09}
Wright S., Larkin J., Law D., Steidel C., Shapley A., Erb D., 2009,
ApJ, 699, 421

\bibitem[Wuyts et al.(2011)]{wuy11}
Wuyts S., et al., 2011,
ApJ, accepted

\end{thebibliography}
\end{document}